\documentstyle[prb,aps,epsf]{revtex}   
\begin{document}
\twocolumn[\hsize\textwidth\columnwidth\hsize\csname %
@twocolumnfalse\endcsname

\title{Persistent Current in the Ferromagnetic Kondo Lattice Model
}
\author{Jun Zang, S. A. Trugman, A. R. Bishop, H. R\"oder}
\address{
    Theoretical Division and Center for Nonlinear Studies, MS B262\\
    Los Alamos National Laboratory, 
    Los Alamos, NM 87545}

\date{\today}
\maketitle
\begin{abstract}
In this paper, we study the zero temperature persistent current in a  
ferromagnetic Kondo lattice model in the strong coupling limit. 
In this model, there are spontaneous
spin textures at some values of 
the external magnetic flux. These spin textures
 contribute a geometric flux, which can induce 
an additional spontaneous 
persistent current. Since this spin texture changes with the
external magnetic flux, we find that there is an anomalous persistent
current in some region of magnetic flux: near $\Phi/\Phi_0=0$
for an even number of electrons and $\Phi/\Phi_0=1/2$ for an odd number
of electrons. 
\end{abstract}
\phantom{.}
]

\narrowtext
\pagebreak

\section{Introduction and Model}
\label{sec:intro}
The Berry phase \cite{berry} plays important roles in transport in mesoscopic
systems. The most well known effect is the persistent current
in quasi-one-dimensional
normal-metal mesoscopic rings threaded with magnetic flux \cite{pc.1,pc.2}.
This purely quantum effect manifests itself through the acquisition
of a phase factor proportional to the magnetic flux which can 
change the boundary condition of the orbital wavefunctions \cite{pc.3}. 
Similarly, when a quantum spin adiabatically follows 
a magnetic field varying  in space or time, the phase of its
wavefunctions acquires an additional contribution -- a geometric
or Berry phase. It was shown by Loss et al. that 
this geometric phase can also induce 
persistent spin and mass currents \cite{pc-s}.

In this paper, we study the persistent current in the
1D ferromagnetic Kondo lattice model (FKL) in the strong coupling limit. 
The FKL model in the strong coupling limit is related to the
double exchange mechanism \cite{note-cmr}, which has been studied recently 
\cite{Kub72,kubo,Mil95,jp95,hzb1,hartmann,zang96}
due to its relevance to the colossal magnetoresistive
Mn-oxides \cite{jonker}. From now on, we will call the
FKL model in the strong coupling limit the double exchange model.
In this double exchange model, 
the ground state  has spin textures
at some values of magnetic flux \cite{kubo,zang96}. 
Since the electronic spins are strongly ferromagnetically coupled
to the local spins,
the spin texture of
the local spin can induce a Berry phase in the electronic wavefunctions.
The geometric flux produced by the local spin texture then induces spontaneous
persistent currents.
In previously studied persistent currents 
induced by geometric flux \cite{pc-s}, 
the geometric
Berry phase is due to adiabatically varying external magnetic field textures. 
In the double exchange model, the geometric Berry phase is due to the
local spin textures,  which depend self-consistently on the 
dynamics of the electrons.
With the change of external magnetic flux, the local spin texture varies.
This in turn will change the geometric phase-induced persistent current.

Here we study the {\it zero} temperature persistent
current of an ideal one-channel double exchange ring threaded with a
magnetic flux $\Phi$. The effects of finite temperature and
disorder (elastic scattering) 
for double exchange systems are similar to those of the conventional
persistent current in normal metal rings \cite{pc-t}, and will not
be considered here.
The FKL model for a $N$-site 1D ring with magnetic 
flux $\Phi$ can be written as
\begin{eqnarray}
H= -t\sum_{ j, \sigma }
    \Bigl( e^{-{i2\pi \Phi\over N\Phi_0}}
    c_{j \sigma}^\dagger
           c_{j+1 \sigma}^{\phantom{\dagger}}+h.c. \Bigr)
           \nonumber \\
   - {J_H\over2} \sum_{i ,  \sigma , \sigma'}
   c_{i \sigma}^\dagger
    ({ \vec \sigma})_{\sigma \sigma'}^{\phantom{\dagger}}
   c_{i \sigma'}^{\phantom{\dagger}}
   \cdot {\vec S}_i ,
\label{eq:ham}
\end{eqnarray}
where $\Phi_0=hc/e$, $\vec{\sigma}$ is the Pauli matrix,
 and $\vec{S}_i$ is the
local spin. The operators 
$c_{i\sigma}^{\phantom{\dagger}}$ 
($c_{i\sigma}^{\dagger}$) annihilate (create) a 
mobile electron with spin $\sigma$. 
For the double exchange model, $J_H/t\gg 1$.
The zero temperature persistent current
is equal to \cite{pc.3} $I=-c\partial E/\partial \phi$. 
It is easy to show that the free energy and persistent current are
periodic in $\Phi/\Phi_0$.
In Hamiltonian (\ref{eq:ham}), we have neglected electron-electron
interactions. In the double exchange model, the {\it on-site} electron-electron
interaction is unimportant; it mainly renormalizes $J_H$.

In Sec.~\ref{sec:classical}, we will study the properties of
spin textures and persistent current in the classical spin limit.
In this limit, many results can be obtained analytically.
In Sec.~\ref{sec:semiclassical}, we study the effects of
quantum fluctuations of order $1/S$ using a set of
trial wavefunctions. In Sec.~\ref{sec:s=1/2}, we study the
spiral ordering and persistent current for $S=1/2$
rings using exact diagonalization. We summarize our
results in Sec.~\ref{sec:sum}.

\begin{figure}[h]
\vskip -0.0cm
 \centerline{
 \epsfxsize=7.5cm \epsfbox{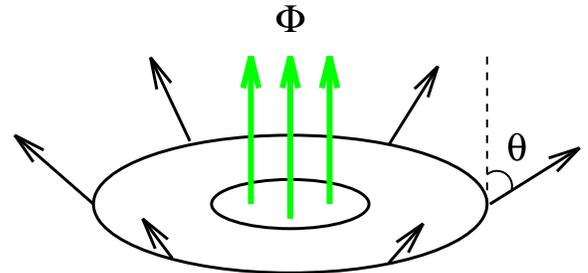}}
\caption{ Local spin  configuration with spiral ordering
${\hat n}_j= (\theta_j,\phi_j)=(\theta,j\phi_0)$. 
\label{fig:sc} }
\end{figure}

\begin{figure}[h]
\vskip -0.0cm
 \centerline{
 \epsfxsize=6.5cm \epsfbox{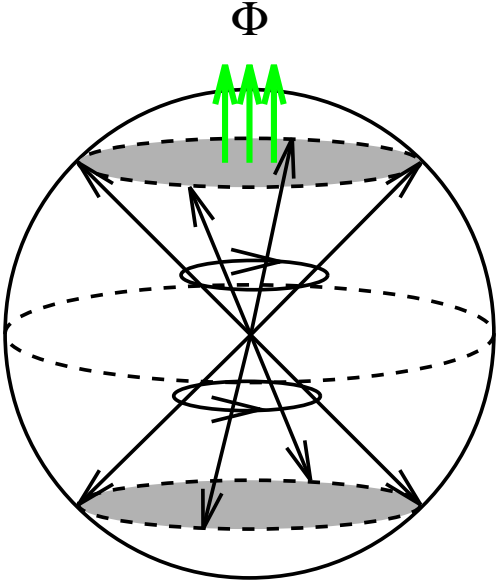}}
\caption{Two equivalent local spin  configurations.
\label{fig:sc2} }
\end{figure}

\section{Classical spin limit $S\rightarrow \infty$}
\label{sec:classical}

The {\it zero} temperature persistent current is most easily calculable
in the classical local spin limit. In this limit, the
electrons are moving in a frozen local spin background.
Hamiltonian (\ref{eq:ham}) is then transformed to a {\it one-body}
Hamiltonian. The ground state of Hamiltonian (\ref{eq:ham})
is the ground state with optimal {\it static} local spin configuration.
Since the original Hamiltonian is translation invariant and
there is an effective short range ferromagnetic coupling between
local spins due to the double exchange mechanism \cite{note-cmr},
we conjecture that the ground state has spiral ordering
\begin{equation}
{\hat n}_j= (\theta_j,\phi_j)=(\theta,j\phi_0), 
\label{eq:sc}
\end{equation}
where $\phi_0=2m\pi/N$ is an integral multiple of
$2\pi/N$ [see Fig.~\ref{fig:sc}].
Note that the maximally polarized ferromagnetic ordering is
an extreme case of this ``spiral ordering'' with $\theta=0$.
Here $\theta$ is a { variational} parameter. We need to 
minimize the total energy $E_{tot}(\theta)$ to find the optimal
$\theta$ for the local spin configurations.

Using the local spin configurations of Eq.(\ref{eq:sc}), 
the Hamiltonian in  $k$-space is
\begin{eqnarray}
\tilde{\cal H}=\left( \begin{array}{lr}
        -{SJ_H\over2}-T_1  &  V_{\delta} 
           \\
        V_{\delta} & {SJ_H\over2}-T_2  
       \end{array}
\right)  \, ,
\label{eq:ham2}
\end{eqnarray}
where $V_{\delta}=\Delta_k\sin\theta$, $\Delta_k=
t\cos(k-\phi_0-{2\pi\Phi\over N\Phi_0})-t\cos(k-{2\pi\Phi\over N\Phi_0})$; 
$T_1=\epsilon^0_k-\Delta_k\cos\theta$ and
$T_2=\epsilon^0_k+\Delta_k\cos\theta$
 are the 
kinetic energies with $\epsilon^0_k=t\cos(k-{2\pi\Phi\over N\Phi_0})
+t\cos(k-\phi_0-{2\pi\Phi\over N\Phi_0})$. From Eq.(\ref{eq:ham2}), we can
show that the electronic energies with the
two spin configurations $\theta \leftrightarrow \pi-\theta$,
$\phi_0 \leftrightarrow -\phi_0$ are degenerate [see Fig.~\ref{fig:sc2}].

The lower eigenenergy of the single electron states are 
\begin{equation}
\epsilon_k = -\epsilon^0_k -\sqrt{({SJ_H\over2})^2+\Delta_k^2-\Delta_kSJ_H\cos\theta}.
\label{eq:eg-cl}
\end{equation}
For $SJ_H \gg |\Delta_k^2|$, 
\begin{equation}
\epsilon_k \simeq -{SJ_H\over2} -\epsilon^0_k+\Delta_k\cos\theta-
\Delta_k^2\sin^2\theta/(SJ_H).
\end{equation}
Note that the eigenenergy is periodic in $\Phi/\Phi_0$.

In the limit $S=\infty$, the many-body wavefunction is just the 
Slater determinant of single particle wavefunctions. The total energy is
\begin{equation}
E_{tot}(\theta)=\sum_{|k|\leq k_f} \epsilon_k .
\end{equation}
At zero flux $\Phi/\Phi_0=0$, the ground state for an even number of electrons
has spiral spin ordering\cite{zang96} $\theta=\pi/2$, i.e.
the spins lie on the equator with zero total magnetization.
For densities { not} close to half filling $n = N$ ($n/N \lesssim 0.8$ 
for $SJ_H \sim 12t$), 
the lowest energy spiral state has a single twist: $\phi_0=2\pi/N$.
Actually, with zero flux, all the spiral states with finite $\theta$
and a single twist have lower energy than the lowest ferromagnetic
state ($\theta=0$).  The energy difference is:
\begin{equation}
E_{tot}(\theta)-E_{tot}(0) \simeq
-\sin^2\theta/(SJ_H)\sum_{|k|\leq k_f} \Delta^2_k 
\end{equation}
for $SJ_H \gg |\Delta_k^2|$.
The results for an odd number of electrons is similar but
shifted by $\Phi/\Phi_0 \rightarrow \Phi/\Phi_0+1/2$.
Here the Berry phase plays an important role to lower the energy
of spiral states: with the spiral spin texture, the
 magnitude of the effective
hopping is reduced, but the effective
hopping matrix element has an additional complex phase,
which can lower the total energy of spiral states for even (odd) number
of electrons at $\Phi/\Phi_0=0$ ($\Phi/\Phi_0=1/2$).

\begin{figure}
\vskip -0.0cm
 \centerline{
 \epsfxsize=8.0cm \epsfbox{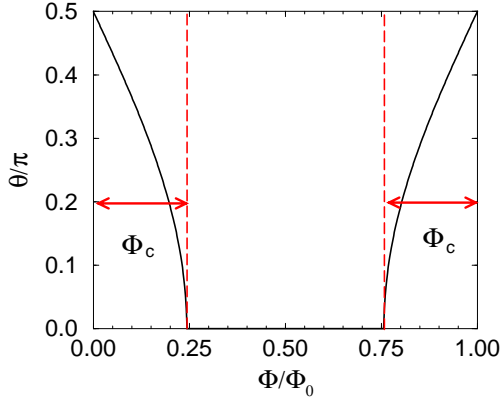}}
\caption{Change of the spiral ordering angle $\theta$ with
magnetic flux $\Phi/\Phi_0$ at $SJ_H=12t$ and $S=\infty$ for a 
$N=450$ ring with
$n=300$.
\label{fig:theta} }
\end{figure}

Let us consider what happens if there is an infinitesimal external
flux ${\Phi\over \Phi_0}\ll 1$. 
For $SJ_H \gg |\Delta_k^2|$,
\begin{eqnarray}
E_{tot}(\theta)-E_{tot}(0) \simeq 
2(1-\cos\theta)t\sin k_f \sin({2\pi\Phi\over N\Phi_0})
\nonumber \\
-\sin^2\theta/(SJ_H)\sum_{|k|\leq k_f} \Delta^2_k .
\label{eq:phi}
\end{eqnarray}
Thus for small external flux ${\Phi\over \Phi_0}\ll 1$, the
spiral angle $\theta$ decreases as
\begin{equation}
\cos\theta \propto \sin({2\pi\Phi\over N\Phi_0}) ,
\end{equation}
and we can expect that as the magnetic flux $\Phi/\Phi_0$ changes, 
the pitch of the spiral ordering 
remains constant (i.e. $\phi_0=2\pi/N$), and
the angle $\theta$ of the spiral ordering decreases continuously.
At some critical
value $\Phi=\Phi_c$, $\theta=0$, i.e. the spins become ferromagnetically
ordered.
In Fig. \ref{fig:theta}, we show the change of spiral ordering angle
$\theta$ with magnetic flux.

\begin{figure}
 \centerline{
 \epsfxsize=8.0cm \epsfbox{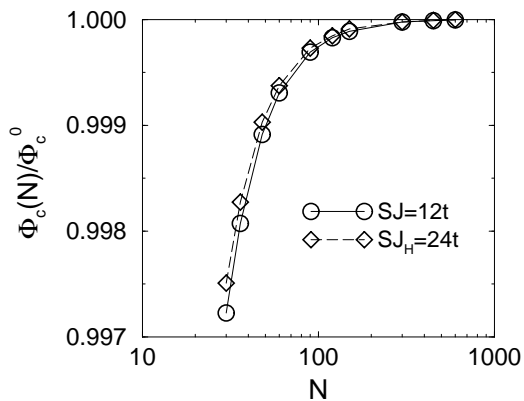}}
\caption{ System size dependence of the critical magnetic flux $\Phi_c$
in the classical spin limit.
$SJ_H=12t, 24t$, density $\nu=2/3$.
\label{fig:f_N} }
\end{figure}

\begin{figure}
\vskip -0.5cm
 \centerline{
 \epsfxsize=8.0cm \epsfbox{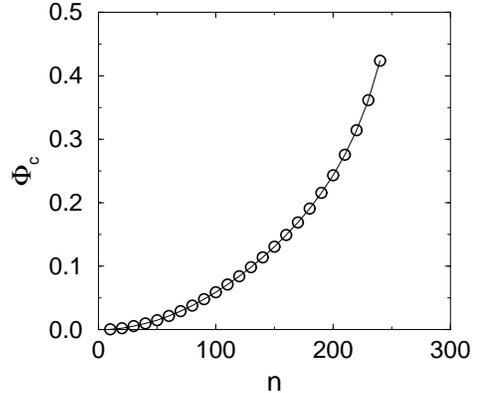}}
\caption{ Density dependence of the critical magnetic flux $\Phi_c$
in the classical spin limit.
$SJ_H=12t$, $N=300$.
\label{fig:f_n} }
\end{figure}

\begin{figure}
\vskip -0.5cm
 \centerline{
 \epsfxsize=8.0cm \epsfbox{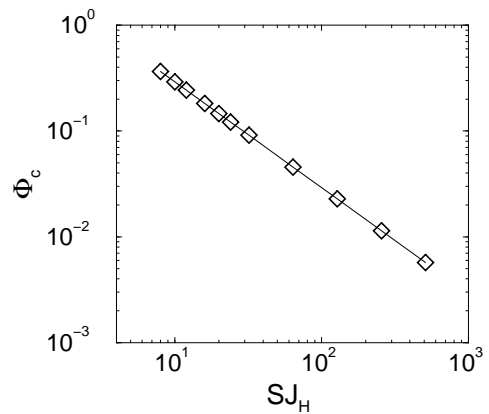}}
\caption{Hund's rule coupling dependence of the critical magnetic flux $\Phi_c$
in the classical spin limit.
$n=200$, $N=300$.
\label{fig:f_j} }
\end{figure}

At fixed density, the critical value of magnetic flux $\Phi_c$
depends only weakly on the size of the ring $N$, as shown
in Fig. \ref{fig:f_N}. From Fig. \ref{fig:f_N}, we can see that
the critical value {\it increases} slightly with increasing system size
$N$ and saturates at $N\lesssim 10^3$. 
As shown in Fig. \ref{fig:f_n}, $\Phi_c$   
depends nearly quadratically on the electron density. From this density
dependence, we can also see that 
there is no particle-hole symmetry in this model.
Since at half filling $n=N$ the ground state is antiferromagnetically
ordered,
our calculations are not valid close to half-filling.
We also show the
$SJ_H$ dependence of $\Phi_c$ in Fig. \ref{fig:f_j},
from which we can conclude that it 
is very close to a power law for
large $SJ_H$: $\Phi_c\sim C/(SJ_H)$.
This $1/(SJ_H)$ dependence is expected from Eq. (\ref{eq:phi}).

\begin{figure}
\vskip -0.0cm
 \centerline{
 \epsfxsize=8.5cm \epsfbox{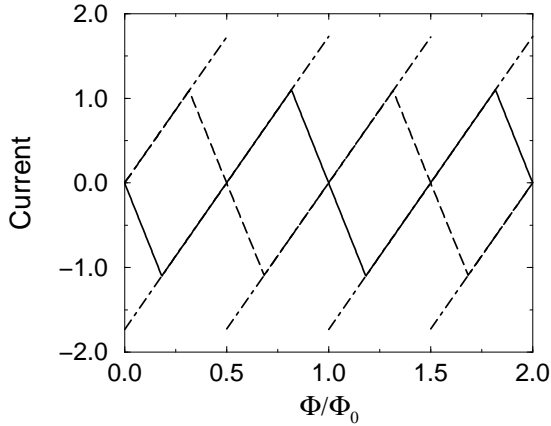}}
\caption{Persistent current in a double exchange ring in the
classical spin limit:
 $N=450$, $n=300$ (solid line), and $n= 301$ (dashed line),
 $SJ_H=16$. The dotted-dashed lines are the persistent currents
 in a conventional mesoscopic metallic ring.
\label{fig:cla} }
\end{figure}

When the local spins are not ferromagnetically ordered, their spin texture
will induce a Berry phase in the electron wavefunctions.
Thus finite $\theta$ will induce a geometric flux, which in turn
induces a spontaneous persistent current in the mesoscopic double exchange
ring. This spontaneous persistent current
will cancel part of the persistent current due to finite
external magnetic flux.
In Fig.(\ref{fig:cla}),  
we show structures of the persistent current 
for even and odd numbers of electrons.
In conventional mesoscopic
systems, the zero temperature persistent current is linear in $\Phi$ for
$k\Phi_0<\Phi<(k+1)\Phi_0$ for an even number of electrons, and
$(k-1/2)\Phi_0<\Phi<(k+1/2)\Phi_0$ for an odd number of electrons.
In a mesoscopic double exchange ring, due to the additional contribution
to the persistent current from spin textures,
the persistent currents are no longer monotonic in the regions
mentioned above.
 Upon ensemble averaging 
of rings with different electron numbers, the period of the
persistent current is changed from $\Phi=\Phi_0$ to $\Phi=\Phi_0/2$.
For double exchange rings, the persistent current
has anomalous regions near both $\Phi=\Phi_0$ and $\Phi=\Phi_0/2$
after ensemble averaging. In the remainder of this paper, we will continue to
discuss even and odd numbers of electrons separately.

\section{semiclassical spin limit $S\gg 1$}
\label{sec:semiclassical}

If the local spin $S$ is large but finite, 
the local spins have quantum fluctuations
of order $1/S$, which can be taken advantage of by electrons
to lower the electronic energy. This is the spin polaronic
effect, which is similar to that in the Hubbard model \cite{szw}. 
Here we still consider the case where
the ground state is either spirally ordered
or ferromagnetically ordered, defined by 
${\hat n}_j= (\theta_j,\phi_j)=(\theta,j\phi_0)$,
as in the classical limit. 

First we consider the eigenstates of the single itinerant electron case.
If one electron is at site $j$, 
the complete spin basis for this site can be chosen as
\begin{eqnarray}
|S+{1\over2},S+{1\over2}\rangle, |S+{1\over2},S-{1\over2}\rangle, 
|S+{1\over2},S-{3\over2}\rangle, \cdots
\nonumber \\
|S-{1\over2},S-{1\over2}\rangle, |S-{1\over2},S-{3\over2}\rangle, 
|S-{1\over2},S-{5\over2}\rangle, \cdots
\label{eq:s}
\end{eqnarray}
with a quantization axis of ${\hat n}_j$.
In the semiclassical limit, these total spin states 
$|S\pm{1\over2},S\pm{1\over2}-k\rangle$ with the local
quantization axis of ${\hat n}_j$ are an appropriate
basis to describe the spin states at the site with
the itinerant electron present. In the classical limit
as studied in the previous section, only
$|S\pm{1\over2},S\pm{1\over2}\rangle$ need be used
to construct the exact wavefunctions. In the semiclassical
limit, the quantum fluctuation corrections will come from
the contributions of $|S\pm{1\over2},S\pm{1\over2}-k\rangle$ ($k>0$).
If we include just the first order quantum fluctuation
effects, we need a { four} trial wavefunction
basis:
\begin{eqnarray}
|\psi_j^{(1)}\rangle &=& |S+{1\over2},S+{1\over2}\rangle_{\displaystyle {\hat n}_j}
^{\phantom{\displaystyle {\hat n}_j}}
 \otimes \{ |S,S\rangle_{\displaystyle {\hat n}_{i\neq j}}
^{\phantom{\displaystyle {\hat n}_{i\neq j} } }      \}
\nonumber \\
|\psi_j^{(2)}\rangle &=& |S-{1\over2},S-{1\over2}\rangle_{\displaystyle {\hat n}_j}
^{\phantom{\displaystyle {\hat n}_j}}
 \otimes \{ |S,S\rangle_{\displaystyle {\hat n}_{i\neq j}}
^{\phantom{\displaystyle {\hat n}_{i\neq j} } }      \}
\nonumber \\
|\psi_j^{(3)}\rangle &=& |S+{1\over2},S-{1\over2}\rangle_{\displaystyle {\hat n}_j}
^{\phantom{\displaystyle {\hat n}_j}}
 \otimes \{ |S,S\rangle_{\displaystyle {\hat n}_{i\neq j}}
^{\phantom{\displaystyle {\hat n}_{i\neq j} } }      \}
\nonumber \\
|\psi_j^{(4)}\rangle &=& |S-{1\over2},S-{3\over2}\rangle_{\displaystyle {\hat n}_j}
^{\phantom{\displaystyle {\hat n}_j}}
 \otimes \{ |S,S\rangle_{\displaystyle {\hat n}_{i\neq j}}
^{\phantom{\displaystyle {\hat n}_{i\neq j} } }      \} ,
\label{eq:trial2}
\end{eqnarray} 
where $|\cdots\rangle_{\displaystyle {\hat n}_{i\neq j}}
^{\phantom{\displaystyle {\hat n}_{i\neq j} } }$ is the spin state
$|\cdots\rangle$ with the local quantization axis ${\hat n}_j$,
which is a generalized coherent state \cite{sp-cs} in the global
quantization axis.
Although we cannot show that the trial wavefunction basis in
Eq.(\ref{eq:trial2}) rigorously describes the $1/S$ corrections
to the results in the classical spin limit,
we believe this will at least describe the qualitative
effects of $1/S$
corrections. Note also that one can include a variational
component in the empty sites wavefunctions such as
$\{ |S,S\rangle_{\displaystyle {\hat n}_{i\neq j}}
^{\phantom{\displaystyle {\hat n}_{i\neq j} } }
+c |S,S-1\rangle_{\displaystyle {\hat n}_{i\neq j}}
^{\phantom{\displaystyle {\hat n}_{i\neq j} } }
\}$,
where $c$ is a variational parameter. But one can show
that these corrections to the eigenenergy
will be higher order in $1/S$ and
thus negligible.

Using the variational basis (\ref{eq:trial2}), 
the Hamiltonian in $k$-space can be
readily written down.  For simplicity, we will only discuss
the large Hund's rule coupling limit $SJ_H =\infty$.  Then we only
need the basis states $|\psi_{j}^{(1)}\rangle$ and 
$|\psi_{j}^{(3)}\rangle$. The Hamiltonian in this variational $2\times2$ 
basis is
\begin{eqnarray}
\tilde{\cal H}=\left( \begin{array}{lr}
        -{SJ_H\over2}-T_1   & \sqrt{1\over 2S+1}V_{\delta}  \\
       \sqrt{1\over 2S+1}V_{\delta}  & -{SJ_H\over2}-T_3 
       \end{array}
\right)
\label{eq:ham3}
\end{eqnarray}
where  $T_3={1\over2S+1}(\epsilon^0_k+\Delta_k\cos\theta)$. 

From the above
trial wavefunctions we can see that quantum fluctuations
make the spins near the electron more polarized 
than the average spiral ordering direction.  Thus
we can think of the quantum fluctuation effect
as a spin polaron effect.
In the limit $J_H\rightarrow\infty$, 
the eigenenergies of the single electron states
are
\begin{eqnarray}
\epsilon_k= -{SJ_H\over 2} - {1\over2} (\epsilon_k^0-\Delta_k\cos\theta+\delta)
\nonumber \\
\nonumber \\
-{1\over2} \sqrt{(\epsilon_k^0-\Delta_k\cos\theta-
\delta)^2+{4\over 2S+1}\Delta^2_k\sin^2\theta} ,
\label{eq:eg-s}
\end{eqnarray}
where $\delta={1\over 2S+1}(\epsilon_k^0+\Delta_k\cos\theta)$.
When $|\epsilon_k^0-\Delta_k\cos\theta-
\delta| \gg |2\Delta_k\sin\theta/|\sqrt{2S+1}$,
\begin{equation}
\epsilon_k= -{SJ_H\over 2} - \epsilon_k^0+\Delta_k\cos\theta
- {\Delta^2_k\sin^2\theta \over 2S
|\epsilon_k^0-\Delta_k\cos\theta-
\delta| } \, .
\label{eq:eg-ss}
\end{equation}

\begin{figure}
\vskip -0.0cm
 \centerline{
 \epsfxsize=8.5cm \epsfbox{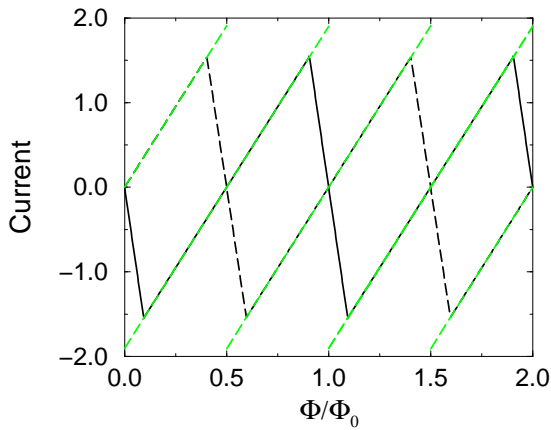}}
\caption{Persistent current in a double exchange ring in the
semiclassical spin limit:
 $N=300$, $n=120$ (solid line), and $n= 121$ (dashed line), $S=5$,
 $J_H=\infty$. The grey long-dashed lines are the persistent currents
 in a conventional mesoscopic metallic ring.
\label{fig:s300} }
\end{figure}

In the semiclassical limit, we can construct {\it approximate}
many-body wavefunctions using Slater determinants of
single particle wavefunctions. However, this construction
is only approximate because we neglect the interaction 
between different spin polarons.
The corrections due to the spin polaron interaction is
$O(n/S^2)$.
Thus this approximation is only valid for the dilute electron limit
and large $S$.
Similar to the finite $J_H$ effects, the quantum fluctuations
induce a spiral ordering instability at $\Phi/\Phi_0=0$ for an even number of
electrons and $\Phi/\Phi_0=1/2$ for an odd number of electrons.
From Eq.(\ref{eq:eg-ss}), 
the lowering of the GS energy with spiral ordering
due to quantum fluctuations is 
$E_{sp}-E_{FM}\propto -|t|/S$. 
Figure (\ref{fig:s300}) show the persistent current
for semiclassical spins.
For low electron densities, the $1/S$ quantum fluctuation effects
are similar to those of finite $SJ_H$ in the classical spin limit.

\section{Persistent current in $S=1/2$ systems --exact diagonalization study}
\label{sec:s=1/2}

\begin{figure}
\vskip -0.0cm
 \centerline{
 \epsfxsize=9.5cm \epsfbox{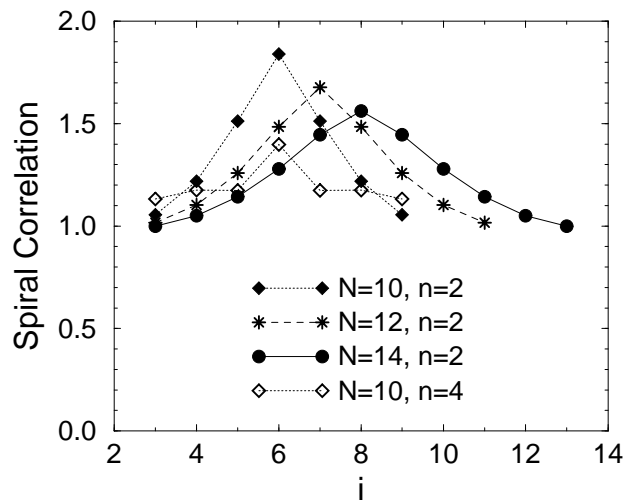}}
\vskip -0.0cm
\caption{Correlation 
$\langle({\vec S}_1\times{\vec S}_2)\cdot({\vec S}_i\times{\vec S}_{i+1})
\rangle$ for $S=1/2$ rings normalized by $(0.25\sin(2\pi/N))^2$. $SJ_H=40t$.
\label{fig:sp} }
\end{figure}

From the results in the classical and semiclassical limits,
we expect that the spiral ordering instability
will be enhanced in the quantum spin limit. 
This instability for general spin $S$ has been demonstrated 
previously \cite{zang96} using 
a many-body trial wavefunction. To confirm that the ground state spin texture
is indeed spiral for a quantum spin, we calculate the exact ground state
for $S=1/2$ rings using exact diagonalization\cite{zang96}.
For an even number of electrons, we find that the
ground state has total spin $S_{total}=0$ for both large
$SJ_H$ and $SJ_H=\infty$ at $\Phi/\Phi_0=0$. With an increase of
external flux, the total spin $S_{total}$ of the ground state
increases, to $S_{total}=S^{max}\equiv (NS+n/2)$ at
$\Phi/\Phi_0=1/2$. For an odd number of electrons, the results
are shifted by $\Phi/\Phi_0\rightarrow \Phi/\Phi_0 + 1/2$.
We speculate that the non-ferromagnetic state with
$S_{total}<S^{max}$ is a sum over global rotations of spiral
states with polarization angle $\theta$ ($0\le\theta<\pi/2$). 
To check this statement numerically, we calculate
the correlation function 
$\langle({\vec S}_1\times{\vec S}_2)\cdot({\vec S}_i\times{\vec S}_{i+1})
\rangle$  for even numbers of electrons at $\Phi/\Phi_0=0$.
Because this ground state with $S_{total}=0$ 
can be thought of as a sum over all global rotations,
we need to look at scalar correlation functions.
The most relevant correlation function
for the spiral ordering is 
$\langle({\vec S}_1\times{\vec S}_2)\cdot({\vec S}_i\times{\vec S}_{i+1})\rangle$.
In Fig.(\ref{fig:sp}), we show the spiral correlation
function
$\langle({\vec S}_1\times{\vec S}_2)\cdot({\vec S}_i\times{\vec S}_{i+1})
\rangle$ for different size rings of spin $S=1/2$.
The correlation is only weakly
dependent on $i$ and has magnitude $\sim (0.25\sin(2\pi/N))^2$.
This correlation function for $S=1/2$ is closer to $(0.25\sin(2\pi/N))^2$ 
as the system size N and/or
electron number n increases.
This clearly demonstrates that the spin ordering is spiral, similar to
what we find in the classical and semiclassical limits.

\begin{figure}
\vskip -0.0cm
 \centerline{
 \epsfxsize=9.5cm \epsfbox{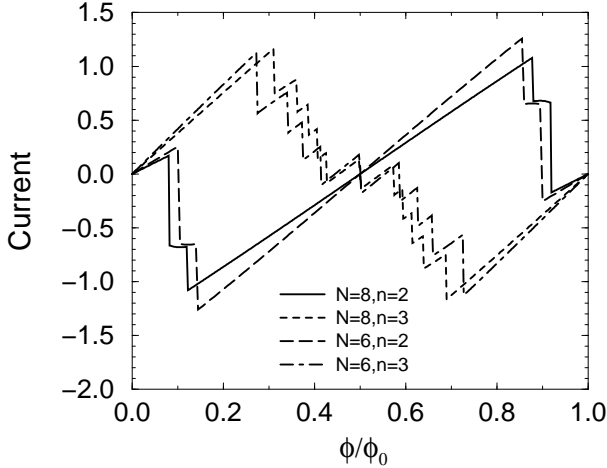}}
\vskip -0.0cm
\caption{Persistent currents in the $S=1/2$ double exchange system.
N=6,8 and  n=2,3. $SJ_H=40t$.
\label{fig:s2} }
\end{figure}

The geometrical flux due to the  local spin textures also contributes 
to the persistent currents in quantum spin systems. 
In Fig. (\ref{fig:s2}),
we show the exact diagonalization results for the persistent current
in $S=1/2$ double exchange rings. 
We can see from Fig. (\ref{fig:s2}) that the width of the anomalous
region is increased compared to that in the classical 
[see Fig. (\ref{fig:cla})] and semiclassical limit [see Fig. (\ref{fig:s300})].
This demonstrates that the possibility of
spiral spin ordering is enhanced in the quantum spin limit,
as expected from the $1/S$ results. Although we do not have enough numerical
results for $\Phi_c$ to study the dependences on system size $N$, 
electron density $n$, and $SJ_H$,  we believe that
these dependences  will be similar to those for classical
local spin double exchange rings, as shown in 
Figs.(\ref{fig:f_N})-(\ref{fig:f_j}).

\section{Conclusion}
\label{sec:sum}

In conclusion, we have studied the novel zero temperature 
persistent current in the 
double exchange system. In this system, there are spontaneous
spin textures at some values of the external magnetic flux. These spin textures
contribute a geometric flux, which can induce additional spontaneous 
persistent current. Since the spin textures vary with the change of
external magnetic flux, there are anomalous persistent
currents in the region near $\Phi/\Phi_0=0$
for an even number of electrons and $\Phi/\Phi_0=1/2$ for an odd number
of electrons. After ensemble averaging of double exchange
rings with different electron numbers, the persistent current
has a period of $\Phi=\Phi_0/2$,
and the anomaly occurs at $\Phi/\Phi_0=0$ and $1/2$. The
effect of disorder (elastic scattering) and finite temperature
for the double exchange systems are similar to those in conventional
normal metal rings \cite{pc-t}.
With the advance of fabrication technology, this fascinating anomalous
persistent current and the magnetic flux dependent
spin texture may become observable.

{\bf Acknowledgements:} 
This work was conducted under the auspices of the US Department of Energy,
and
supported (in part) by funds provided by the University of California for
the conduct of discretionary research by Los Alamos National Laboratory.

\end{document}